%% file: paper.tex
\documentclass{article}

\usepackage[nonatbib,preprint]{neurips_2023}

\usepackage[utf8]{inputenc} 
\usepackage[T1]{fontenc}    
\usepackage{hyperref}       
\usepackage{url}            
\usepackage{booktabs}       
\usepackage{amsfonts}       
\usepackage{nicefrac}       
\usepackage{microtype}      
\usepackage{xcolor}         
\usepackage{mathtools}
\usepackage{amsmath}
\usepackage{cite}
\usepackage{algorithm}
\usepackage{algpseudocode}
\usepackage{multirow}
\usepackage[bottom]{footmisc}

\title{A Performance Model for Warp Specialization Kernels}

\author{%
Zhengyang Liu\thanks{This work was done while working at NVIDIA.}\\
University of Utah\\
\texttt{liuz@cs.utah.edu} \\
\And
Vinod Grover\thanks{Address all correspondence to vgrover@nvidia.com }\\
NVIDIA\\
\texttt{vgrover@nvidia.com} \\
}

\begin{document}

\maketitle

\begin{abstract}

This paper presents a performance model tailored for
warp specialization kernels, focusing on factors such as warp size,
tilling size, input matrix size, memory bandwidth, and thread
divergence. Our model offers accurate predictions of execution time by
leveraging differential equations validated through simulations
and experiments. The insights gained from this model not only
enhance our understanding of warp specialization techniques but also
have practical implications for optimizing GPU-accelerated
applications through compiler optimizations, kernel parameter tuning,
and algorithm design.

\end{abstract}

\input{background}

\input{model}
\input{solving}
\input{evaluation}

\input{conclusion}

\bibliography{reference}{}
\bibliographystyle{plain}

\end{document}

%% file: background.tex
\section{Background}

\subsection{General Matrix Multiplication}

A General Matrix Multiply kernel (GeMM) is a specialized computational
routine for performing matrix multiplication, often used in numerical
computing and various applications like machine learning and
scientific computations.
A GeMM kernel takes two input matrices, $\mathbf{A}$ and $\mathbf{B}$,
and produces an output matrix $\mathbf{C}$.
$$
\mathbf{C = A B}
$$
The matrix $\mathbf{A}$ has a dimension of $ M \times K $,
the matrix $\mathbf{B}$ has a dimension of $ K \times N$, and the
output matrix $\mathbf{C}$ has a dimension of $M \times N$.

In the context of GeMM, \emph{tiling} refers to a technique used to
improve the data locality by dividing the input matrices into smaller,
regularly-sized sub-matrices and performing the multiplication on
these smaller sub-matrices, instead of working on the the original
matrices.

\begin{figure}
\centering
\includegraphics[width=0.5\textwidth]{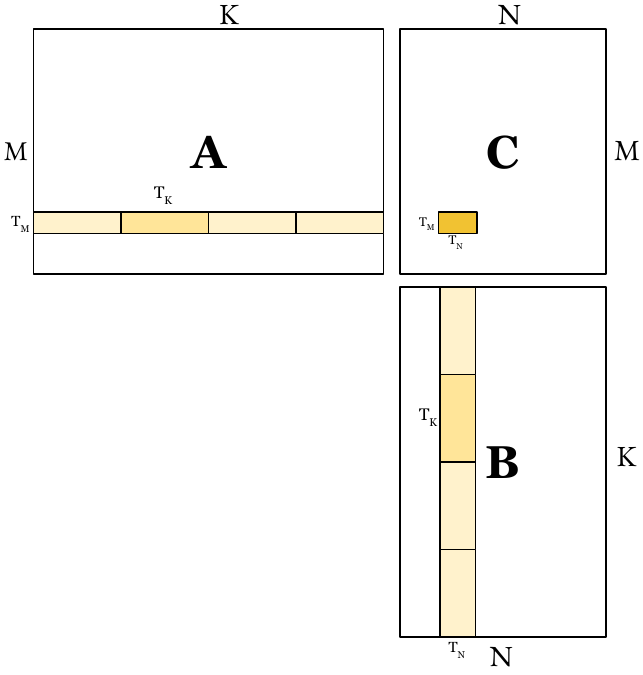}
\caption{GeMM with tiling}
\label{fig:tiling}
\end{figure}

Figure~\ref{fig:tiling} shows an example of tiling.
In this work, we represent a tiling configuration with a tuple of
three non-zero integers, denoted as $(T_M, T_N, T_K)$.
In a tiling configuration $(T_M, T_N, T_K)$, the first input matrix is
split into tiles with dimension $T_M \times T_K$, and the second input is
split into tiles with dimension $T_K \times T_N$. The output matrix is
split into tiles with dimension $T_M \times T_N$.

\subsection{Warp Specialization}

GeMM kernels are often implemented with sequential code that loads
the data from global memory, followed by the matrix multiplication,
and finally stores the result back.
A drawback is that, the computation and the data accesses are
synchronized, which leads to the idle time in either the Arithmetic
Logic Unit (ALU) or the Direct Memory Access unit (DMA).
To avoid the idles, the GeMM kernel can be pipelined; the DMA engine
can start to load the next iteration of data while the compute units
are performing the multiplication.
%
Such pipelined kernels requires a finer-grained synchronizations
between the ALU and the DMA.
Usually, the synchronization is implemented with a producer-consumer
model, where the ALUs are the consumers, and the DMA units are
producers.


%
%
Warp specialization kernels \cite{bauer2014singe} are kernels that
dynamically branches based on each thread's ID, and each warp is
specialized to perform a different task.
In the context of GeMM, each warp is specialized to perform a
different stage of the matrix multiplication.
With per-warp specialization, when carefully scheduled, the access of
data and computation can be executed in parallel, while avoids the
performance degradation from in-warp branching.

\subsection{The Warp Specialized GeMM Kernel in CUTLASS}

In this work, we focus on the warp specialized GeMM kernel implemented
in the NVIDIA CUTLASS library (GeMM-WS) \cite{CUTLASS}.
The GeMM-WS kernel loads the input matrices from global memory,
multiplies them, and stores the result back.
A pseudocode of the GeMM-WS kernel is shown in Algorithm~\ref{alg:gemmws}.
In the GeMM-WS, each warp is specialized by its functionality.
There are two kinds of specialized warps: the \texttt{DMA} warps which
loads the input tiles from global to the shared memory, and the
\texttt{MATH} warps which loads the input tile from shared memory to
registers, performs the GeMM computation then stores the output tile
back to the global memory.
We call the process where the \texttt{DMA} warp loads the data, the
\textit{prologue}, and the process where the \texttt{MATH} warp writes
the results back, the \textit{epilogue}. A \texttt{MATH} is associate
with one or two \texttt{DMA} warps;
In one \texttt{MATH}, one \texttt{DMA} configuration, as GeMM is a
binary operation, one DMA warps are associated with each \texttt{MATH}
warp, each \texttt{DMA} tile loads two input tiles, while in one
\texttt{MATH}, two \texttt{DMA} configuration, two \texttt{DMA} warps
are associated with each \texttt{MATH} warp, and each of \texttt{DMA}
warps loads one input tile.

GeMM-WS allocates a circular buffer in the shared memory in
initialization.
When a \texttt{DMA} warps loads a tile from global
memory, it \textit{pushes} the tile into the buffer. When a
\texttt{MATH} warp completes the computation, it \textit{pop} the tile
out of the buffer.
At the time when the buffer is full, the \texttt{DMA} warps must wait
as there is no spaces left in the buffer. Similarly, when the buffer
is empty, the \texttt{MATH} warps must wait as there is no tiles
available for compute.

GeMM-WS enforces two synchronizations.

\begin{enumerate}
\item{The kernel implements a wait-signal semaphore that
enforces the happen-before relation between the \texttt{DMA} warps and
the \texttt{MATH} warps.}
\item{In any time, there must be at most one active \texttt{MATH}
warp, as multiple active \texttt{MATH} warps may cause load imbalance,
which degrades the overall performance.}
\end{enumerate}

\begin{algorithm}
\caption{Pseudocode of the GeMM-WS kernel}
\label{alg:gemmws}
\begin{algorithmic}[1]
\State{initialize()}
\For{$k \gets 1$ to $\lceil \frac{K}{T_K} \rceil$}
  \If {current warp is a \texttt{DMA} warp}
    \State wait()
    \State \texttt{lhs} $\gets$ load\_input\_tiles($i$, $k$)
    \State \texttt{rhs} $\gets$ load\_input\_tiles($k$, $j$)
    \State signal()
  \ElsIf {current warp is a \texttt{MATH} warp}
    \State wait()
    \State \texttt{result} $\gets$ compute(\texttt{lhs}, \texttt{rhs})
    \State signal()
    \State write\_output\_tile(\texttt{result})
  \EndIf
\EndFor
\end{algorithmic}
\end{algorithm}

%% file: model.tex
\section{Performance Model}
\label{model}

We created a performance model for predicting the execution behavior
of a warp specialization kernel.
The smallest unit of scheduling in our model is a warp.
We call a \texttt{DMA} warp as a producer and a \texttt{MATH} warp as
a consumer.
We assume only one \texttt{MATH} warp or one \texttt{DMA} warp can be
active at any time.
%

\subsection{High Level Performance Analysis}

In this section, we analyze the performance of the GeMM-WS kernel at a
high level.

\subsubsection{Wave}

In the GeMM-WS kernel, the computation of each \textit{output} tile is
mapped to a cooperative thread array (CTA).
The number of CTA allocated during a kernel launch is bounded by
either number of output tiles required, or the number of streaming
multiprocessors (SMs) in the system, whichever is smaller.
If the number of output tiles is less or equal than the numbers SMs in
the system,
we can distribute the computation of every output tile to a single SM.
In this case, each SM has at most one tile to compute, so the
computation can finish within a single \emph{wave}.
Otherwise, the computation cannot finish within one wave; for
example, when there are 256 output tiles to compute, the tiles are
assigned evenly to 84 SM, which as a result, 80 of SM are assigned
with three output tiles, while the rest four SM are assigned with four
output tiles.
Overall, the workloads takes four waves to finish.

In general, the number of waves required to finish the GeMM is

$$ W = \lceil \frac{\text{Number of Output Tiles}}{\text{Number of SMs}} \rceil $$

The number of output tiles can be calculated by multiplying the tiles
counts in $M$ and $N$ dimensions.

$$ \text{Number of Output Tiles} = \lceil \frac{M}{T_M} \rceil \times \lceil \frac{N}{T_N} \rceil $$

The overall execution time of the GeMM-WS kernel is the sum of the
execution time of each wave, offset by the time when the first wave
starts.

$$ \text{Overall Execution Time} = \sum_{i=1}^{W} \text{Wave Execution Time} + T_\text{INIT} $$

In this work, we assume the execution time of each wave is the same, that is,
$T_M$ and $T_N$ are divisible by $M$ and $N$, respectively.

$$ \text{Overall Execution Time} =  \text{Wave Execution Time} \times W + T_\text{INIT} $$

\subsubsection{Stage}

As a result of the tiling on the $K$ dimension on the input tile, the
computation for output tile is split into \emph{stage}.
In each stage, the kernel loads and multiplies an input tile with
dimension $(T_M, T_K)$ from the first input matrix and an input tile with
dimension $(T_K, T_N)$ from the second input matrix, then accumulate
the results to the output tile.
In a tiling configuration $(T_M, T_N, T_K)$,
an output tile requires $S$ stages to finish.

$$S = \lceil \frac{K}{T_K} \rceil$$

For a trivial synchronous GeMM kernel without any overlap,
the overall execution time of a wave is the sum of the execution time of each stage.

$$ \text{Wave Execution Time} = \sum_{i=1}^{S} \text{Stage Execution Time}(i) $$

Similarly, if we assume the execution time of each stage is the same, that is,
$T_k$ is divisible by $K$.

$$ \text{Wave Execution Time} = \text{Stage Execution Time} \times S $$

The overall execution time of a synchronous GeMM kernel is

$$ \text{Overall Execution Time} = (\text{Stage Execution Time}
   \times S ) \times W + T_{\text{INIT}} $$

\subsection{Event Analysis in a Wave}

As in the GeMM-WS kernel, the computation and loading are overlapped,
we need to further analyze the events within a wave to predict the
execution time of a wave.
This section focuses on the events in a wave.

We define three kinds of events in a wave.
\begin{enumerate}
    \item A \texttt{DMA} warp starts to load input tile $a_i$ from
    input matrix $\mathbf{A}$ to shared memory in stage $i$. The start
    times of the loads are denoted as $S_a(i)$. \footnote{Index start from 1 in this work.}
    \item A \texttt{DMA} warp starts to load input tile $b_i$ from
    input matrix $\mathbf{B}$ to shared memory in stage $i$. The start
    times of the loads are denoted as $S_b(i)$.
    \item A \texttt{MATH} warp starts to multiply the input tiles and
    accumulate the result to the output tile in stage $i$. The start
    time of the multiplication in are denoted as $S_m(i)$.
\end{enumerate}

We make the following assumptions.

\begin{enumerate}
    \item The overhead of querying the semaphore value is constant.
    \item The loading time from global memory is directly proportional
    to the number of bytes involved.
    \item We ignore the time spent storing the result back to global
    memory as it is always overlap with other \texttt{MATH} or
    \texttt{DMA} warps except for the final iteration thus we believe
    it is negligible for the final execution time.
\end{enumerate}

Our performance model introduces three differential equations that
model the relation between $S_a$, $S_b$, and $S_m$.
We denote the time needed for computing a tile as
$\texttt{T}_{\text{MATH}}$, and the time needed for loading an input
tile from $A$ and $B$ to shared memory as
$\texttt{T}_{\text{LOAD-A}}$ and $\texttt{T}_{\text{LOAD-B}}$.
Those times are linear to the number of bytes involved in the
operation, offset by the startup latency of the \texttt{MATH} and
\texttt{DMA} warps, respectively.

$$ \texttt{T}_{\text{MATH}} = \frac{T_M \cdot T_N \cdot T_K}{\texttt{ComputeThrouput}} + \texttt{ComputeStartupLatency} $$

$$ \texttt{T}_{\text{LOAD-A}} = \frac{T_M \cdot T_K }{\texttt{LoadThrouput}} + \texttt{LoadStartupLatency} $$

$$ \texttt{T}_{\text{LOAD-B}} = \frac{T_K \cdot T_N }{\texttt{LoadThrouput}} + \texttt{LoadStartupLatency} $$

We assume the shared memory can hold at most $M$ tiles at a time, $M > 2$.

\begin{equation} \label{eq:produce_a}
    S_a(i) = \begin{cases*}
                         0                           & if $i = 1 $ \\
                         \max\Bigl(S_b(i-1) + T_{\text{LOAD-B}}, S_m (i - M) + T_{\text{MATH}}\Bigr)  & otherwise
             \end{cases*}
\end{equation}

Equation~\ref{eq:produce_a} denotes the time when a DMA warp starts to
load input tile $a_i$ from input matrix $\mathbf{A}$ to shared memory.
The initial input tile $a_0$ is loaded at time 0.
Otherwise, the start of loading the $i$-th input tile $a_i$ occurs either
after completing the loading the input tile $b_{i-1}$ of previous stage or
upon the \texttt{MATH} warp releasing memory when the buffer reaches full capacity.
As the buffer is circular, as soon as \texttt{MATH} warp releases the
input tile $a_{i-M}$ and $b_{i-M}$, the load of the $i$-th input tiles can start.

\begin{equation} \label{eq:produce_b}
    S_b(i) = \max\Bigl(S_a(i) + T_{\text{LOAD-A}}, S_m (i - M) + T_{\text{MATH}}\Bigr)
\end{equation}

Equation~\ref{eq:produce_b} denotes the time when a DMA warp starts to
load input tile $b_i$ from input matrix $\mathbf{B}$ to shared memory.
Correspondingly, the loading time for input tile $b_i$ is determined
either by the completion of loading the input tile $a_i$ from the same
stage or by waiting until the \texttt{MATH} warp releases memory when
the buffer reaches its full capacity.

\begin{equation} \label{eq:consume}
    S_m(i) = \max\Bigl(S_m(i-1) + T_{\text{MATH}}, S_b (i) + T_{\text{LOAD-B}}\Bigr)
\end{equation}

Equation~\ref{eq:consume} denotes the time when a \texttt{MATH} warp
starts to multiply the input tiles.
When the input tiles are ready, the \texttt{MATH} warp starts to
multiply immediately after the \texttt{MATH} warp finishes
the computation of previous stage.
Otherwise, the \texttt{MATH} warp waits until the input tiles are
ready.

Our model uses the \emph{max} operator to decide if the event is
scheduled to execute immediately or to wait for the resource to be
available. This is correct because whether the event is scheduled to
execute or wait is determined by the release time of the resource.
Specifically, if the event is set for immediate execution, the release
time consistently precedes the scheduled time of the event.
Conversely, if the event is intended to wait, the scheduled time of
the event exceeds the release time of the resource.

As the epilogue is not overlapped with other events for the final stage,
we need to add the time for the last epilogue to the wave execution time.
Therefore, the execution time of a wave is the time when the last stage's
\texttt{MATH} warp finishes the computation.

$$ \text{Wave Execution Time} = S_m(S) + T_{\text{EPILOGUE}} $$

The overall execution time of the GeMM-WS kernel is

$$ \text{Overall Execution Time} =  \Bigl(S_m(S)+T_{\text{EPILOGUE}}\Bigr) \times W + T_{\text{INIT}} $$

%% file: solving.tex
\section{Finding the Optimal Configuration}
\label{sec:solving}

We developed an execution simulation tool and an optimal solution solver.
This section describes the implementation of the two tools and how we
cross-validated the results from the two tools.

\subsection{Execution Simulation}

We developed a simulation tool for emulating the GeMM-WS kernel
execution with our performance model.
The tool takes the input matrix sizes, tile configuration, and
machine configurations as inputs. It outputs the timestamp of every
event in the kernel launch.
The tool allows us to understand the impact of different tile and
machine configurations on the kernel's performance.

The tool initializes the timestamp of the first event of each
wave to zero. Then it iterates through all the stages and updates the
timestamp of each event in the stage. The tool returns the
timestamp of the last event of the last wave as the overall execution
time.
The tool is implemented in Python. It takes less than one second
to simulate the execution of $M = 1024, N = 1024, K = 1024$ input size.

\subsection{Finding optimal solutions with an SMT solver}

A limitation of the simulation tool is that it can only predict the
execution time of a given configuration.
Therefore, we developed a tool based on SMT solver to
find the optimal configuration for a given input matrix size.

We encoded our performance model in the tool. Our encoding is
carefully engineered so that Z3 can return the results efficiently.
To allow solving with variable circular buffer size, the encoding must
allows indexing into $S_a$, $S_b$, $S_m$ with variable indexes, and for this
reason, we encode the problem using the Array theory~\cite{array}.
We use the integer as the type of $S_a$, $S_b$, $S_m$ as integers are
more efficient to solve compared with reals or floating points in Z3.

We created two optimization goals. The first goal is to minimize the overall execution time.

$$ \text{minimize} ~ \Biggl( \Bigl(S_m(S)+T_{\text{EPILOGUE}}\Bigr) \times W + T_{\text{INIT}} \Biggr)$$

Our second optimization goal is to minimize the total waiting time of all the \texttt{MATH} warps.
The total waiting time of all the \texttt{MATH} warps is the sum of the waiting time of all the \texttt{MATH} warps in all the stages.

$$ \text{minimize} ~ \Bigl(W \times \sum_{i=1}^{S}  \text{Wait}(i)\Bigr)$$

For the initial stage, the waiting time of a \texttt{MATH} warp is the
finish time of loading the input tile from matrix $\mathbf{B}$ to shared
memory, for the rest of the stages, the waiting time of a
\texttt{MATH} warp is the difference between the finish time of the
\texttt{DMA} warp in previous stage and the start time of the
\texttt{MATH} warp in current stage.

$$ \text{Wait}(i) = \begin{cases*}
    S_b(1) + T_{\text{LOAD-B}} & if $i = 1$ \\
    S_m(i) - (S_m(i - 1) + T_{\text{MATH}}) & if $i > 1$
\end{cases*}$$


\subsection{Cross Validating the Simulator and the Solver}

To verify the correctness of output from both the simulator and the
solver, we compared the execution time predicted by the simulator and
the solver. We executed both the simulator and the solver on different
$M$, $N$ and $K$ as well as different machine configurations, and
compared the execution time predicted by the simulator and the solver.
We set $M$ and $N$ and $K$ to be values between 32 and 1024, with a
step size of 32. We tested different machine configurations as well,
including the fake scenarios where the DMA speed is significantly
faster or slower than the ALU, and the scenarios where the DMA is on
par with the ALU. The execution time predicted by the simulator and
the solver are identical for all the configurations we tested.

%% file: evaluation.tex
\section{Evaluation}

This section evaluates our performance model, we begin by showing the
synthesis time of our optimizer, and then we compare the execution
time predicted by our model with the real-world execution time of the
kernel.

\subsection{Solving Time}

We ran our solver on various machine configurations and input matrix
sizes, and we found that the solver can find the optimal solution
within a few minutes in most cases. Our solver is implemented in
Python and uses the Z3 version 4.12.2. We ran the solver on a machine
with an Intel Xeon 6112R processor and 64 GB of RAM.
Table \ref{tab:solving} shows the solving time of our solver on
different problem sizes.

\begin{table}[ht]
    \centering
    \small
    \begin{tabular}{c c c | c c c | c}
    \toprule
    \multicolumn{3}{c|}{Problem size} & \multicolumn{3}{c|}{Optimal
    tilling} & Solving time (s) \\
    \midrule
    M & N & K & $T_M$ & $T_N$ & $T_K$ & \\
    \midrule
    1024 & 1024 & 1024 & 128 & 128 & 128 & 0.374 \\
    1024 & 512  & 1024 & 128 & 64  & 128 & 0.556 \\
    1024 & 1024 & 512  & 128 & 128 & 128 & 0.315 \\
    1024 & 512  & 512  & 128 & 64  & 128 & 0.324 \\
    512  & 512  & 512  & 64  & 64  & 128 & 0.210 \\
    \bottomrule
    \end{tabular}
    \caption{Solving time}
\label{tab:solving}
\end{table}

\subsection{Prediction Accuracy}

In this section, we demonstrate how we estimate the parameters in our
performance model with real world measurements.
Then, we use the estimated parameters to predict the execution time of
the GeMM-WS kernel with our performance model.
Finally, we compare the predicted execution time with the real-world
execution time of the kernel.

We ran the GeMM-WS kernel on an NVIDIA A6000 GPU.
We removed the redundant initialization code and the branches that
never executed to reduce the irrelevant factors in the performance
measurements. Each measurement is the average of 10000 runs.

\subsubsection{Measure the Kernel Launch Time}

We measured the kernel launch time of the GeMM-WS kernel,
$T_{\text{INIT}}$, by running the kernel with an empty function body.
The average kernel launch time is $1.680$ us, with a standard
deviation of $0.027$ us. We subtract the kernel launch time from the
execution time of the kernel in the measurement data for the rest of
the evaluation.

\subsubsection{Measure the Epilogue Time}

We measured the execution time of a single epilogue execution,
$T_{\text{EPILOGUE}}$, by running the kernel with a single stage, and
the stage is specialized to perform the epilogue only.
The average epilogue time is $1.543$ us, with a standard deviation of
$0.174$ us.

\subsubsection{Estimate the Throughput and Startup Latency of the
\texttt{DMA} and \texttt{MATH} Warps}

We measured the execution time of loading two input tiles of different
sizes from matrix $\mathbf{A}$, $T_{\text{LOAD-A}}$. With those
numbers we can estimate the startup latency and the throughput of the
\texttt{DMA} warps. The throughput of the \texttt{DMA} warps can be
calculated as:

\begin{equation}
    T_{\texttt{LoadThroughput}} = \frac{T_M' \times T_K' - T_M'' \times T_K'' }{T_{\texttt{LOAD-A}}' - T_{\texttt{LOAD-A}}''}
\end{equation}

where $T_M'$ and $T_K'$ are the tile dimension of the first
measurement, and $T_M''$ and $T_K''$ are the tile dimension of the
second measurement, and $T_{\texttt{LOAD-A}}'$ and
$T_{\texttt{LOAD-A}}''$ are the execution time of the first and second
measurement, respectively. The startup latency of the \texttt{DMA}
warps can be calculated as:

\begin{equation}
    T_{\texttt{LoadStartupLatency}} = T_{\text{LOAD-A}}' - \frac{T_M' \times T_N'}{T_{\texttt{LoadThroughput}}}
\end{equation}

We performed two measurement, one with $T_M' = T_K' = 64$ and the
other with $T_M'' = T_K'' = 128$
%

From our measurement, the DMA throughput is approximately 152.96 MB/s,
while its startup latency is $0.770$ us.

Similarly, we measured the execution time of multiplying two pairs of
input tiles of different sizes. With those numbers we can estimate the
startup latency and the throughput of the \texttt{MATH} warps. The
throughput of the \texttt{MATH} warps can be calculated as:

\begin{equation}
    T_{\texttt{ComputeThroughput}} = \frac{T_M' \times T_N' \times T_K' - T_M'' \times T_N'' \times T_K'' }{T_{\texttt{MATH}}' - T_{\texttt{MATH}}''}
\end{equation}

The startup latency of the \texttt{MATH} warps can be calculated as:

\begin{equation}
    T_{\texttt{ComputeStartupLatency}} = T_{\text{MATH}}' - \frac{T_M' \times T_N' \times T_K'}{T_{\texttt{MathThroughput}}}
\end{equation}

The ALU throughput is approximately 24.61 GB/s, its startup latency is
negligible.

\subsubsection{Estimate the Execution Time of the GeMM-WS Kernel}

We measured the execution time of the GeMM-WS kernel with different
input matrix sizes and tile configurations. For the input size, we
chose every combination of $M$, $N$ and $K$, ensuring that each of
these values is a multiple of 128, up to a maximum of 1024. For the
tile configuration, we chose every combination of $T_M$, $T_N$ and
$T_K$, each $T_M$, $T_N$ and $T_K$ is either 64 or 128.

We then predicted the overall execution time on the same combinations
of input and tiling with our performance model. We compared the
execution time predicted by our model with the execution time measured
from the real-world execution.
For all 2048 different input and tiling configurations we have
evaluated, our model predicts the execution time with an average error
of 4.5\%.
The maximum error is 21.5\%.

Table~\ref{tab:prediction} shows the prediction accuracy of our model
for a subset of the input and tiling configurations.

\begin{table}[h]
\centering
\small
\begin{tabular}{c c c | c c c | c c | r}
\toprule
\multicolumn{3}{c|}{Problem Size} & \multicolumn{3}{c|}{Tilling} &
\multicolumn{2}{c|}{Execute time (s)} & \\
\midrule
M & N & K & $T_M$ & $T_N$ & $T_K$ & Predicted & Measured &
\multicolumn{1}{c}{Error} \\
\midrule

256  & 256  & 256  & 128 & 128 & 64 & 0.00879698 & 0.008188 & 6.92\%
\\ 
256  & 256  & 256  & 128 & 64  & 64 & 0.00837396 & 0.008188 & 2.22\%
\\ 
256  & 256  & 512  & 128 & 128 & 64 & 0.01321139 & 0.012708 & 3.81\%
\\ 
256  & 256  & 512  & 128 & 64  & 64 & 0.0126495  & 0.012708 & -0.46\%
\\ 
256  & 256  & 1024 & 128 & 128 & 64 & 0.02193814 & 0.021748 & 0.87\%
\\ 
256  & 256  & 1024 & 128 & 64  & 64 & 0.02132557 & 0.021748 & -1.98\%
\\ 
256  & 512  & 256  & 128 & 128 & 64 & 0.00894882 & 0.008188 & 8.50\%
\\ 
256  & 512  & 256  & 128 & 64  & 64 & 0.00838637 & 0.008188 & 2.37\%
\\ 
256  & 512  & 512  & 128 & 128 & 64 & 0.01331758 & 0.012708 & 4.58\%
\\ 
256  & 512  & 512  & 128 & 64  & 64 & 0.0127373  & 0.012708 & 0.23\%
\\ 
256  & 512  & 1024 & 128 & 128 & 64 & 0.02233828 & 0.021748 & 2.64\%
\\ 
256  & 512  & 1024 & 128 & 64  & 64 & 0.02136722 & 0.021748 & -1.78\%
\\ 
256  & 1024 & 256  & 128 & 128 & 64 & 0.00904743 & 0.008188 & 9.50\%
\\ 
256  & 1024 & 256  & 128 & 64  & 64 & 0.00860939 & 0.008188 & 4.89\%
\\ 
256  & 1024 & 512  & 128 & 128 & 64 & 0.01349714 & 0.012708 & 5.85\%
\\ 
256  & 1024 & 512  & 128 & 64  & 64 & 0.01289372 & 0.012708 & 1.44\%
\\ 
256  & 1024 & 1024 & 128 & 128 & 64 & 0.02249229 & 0.021748 & 3.31\%
\\ 
256  & 1024 & 1024 & 128 & 64  & 64 & 0.02166768 & 0.021748 & -0.37\%
\\ 
512  & 512  & 256  & 128 & 128 & 64 & 0.0090228  & 0.008188 & 9.25\%
\\ 
512  & 512  & 256  & 128 & 64  & 64 & 0.00871077 & 0.008188 & 6.00\%
\\ 
512  & 512  & 512  & 128 & 128 & 64 & 0.01357635 & 0.012708 & 6.40\%
\\ 
512  & 512  & 512  & 128 & 64  & 64 & 0.01298575 & 0.012708 & 2.14\%
\\ 
512  & 512  & 1024 & 128 & 128 & 64 & 0.02248066 & 0.021748 & 3.26\%
\\ 
512  & 512  & 1024 & 128 & 64  & 64 & 0.0216738  & 0.021748 & -0.34\%
\\ 
512  & 1024 & 256  & 128 & 128 & 64 & 0.00920297 & 0.008188 & 11.03\%
\\ 
512  & 1024 & 256  & 128 & 64  & 64 & 0.00897629 & 0.008188 & 8.78\%
\\ 
512  & 1024 & 512  & 128 & 128 & 64 & 0.01366439 & 0.012708 & 7.00\%
\\ 
512  & 1024 & 512  & 128 & 64  & 64 & 0.01339277 & 0.012708 & 5.11\%
\\ 
512  & 1024 & 1024 & 128 & 128 & 64 & 0.02269439 & 0.021748 & 4.17\%
\\ 
512  & 1024 & 1024 & 128 & 64  & 64 & 0.02226608 & 0.021748 & 2.33\%
\\ 
1024 & 1024 & 256  & 128 & 128 & 64 & 0.00992178 & 0.008188 & 17.47\%
\\ 
1024 & 1024 & 256  & 128 & 64  & 64 & 0.0149218  & 0.014696 & 1.51\%
\\ 
1024 & 1024 & 512  & 128 & 128 & 64 & 0.01476866 & 0.012708 & 13.95\%
\\ 
1024 & 1024 & 512  & 128 & 64  & 64 & 0.02366335 & 0.023736 & -0.31\%
\\ 
1024 & 1024 & 1024 & 128 & 128 & 64 & 0.02514989 & 0.021748 & 13.53\%
\\ 
1024 & 1024 & 1024 & 128 & 64  & 64 & 0.04187472 & 0.041816 & 0.14\%
\\ \bottomrule

\end{tabular}
\caption{Prediction accuracy for selected input sizes and tilings}
\label{tab:prediction}
\end{table}

%% file: conclusion.tex
\section{Conclusion}
\label{sec:conclusion}

In this paper, we studied the runtime performance of the GeMM-WS
kernel.
We presented a performance model for predicting the
execution time of a warp specialization kernel.
Based on the performance model, we developed an execution simulation
tool and an optimal solution solver.
We used the two tools to find the optimal configuration for the GeMM-WS
kernel.
Our solver can find the optimal configuration within a few minutes in
most cases.
We also demonstrated that our performance model can predict the
execution time of the GeMM-WS kernel with high accuracy.

%% file: paper.bbl
\begin{thebibliography}{1}

\bibitem{bauer2014singe}
Michael Bauer, Sean Treichler, and Alex Aiken.
\newblock Singe: Leveraging warp specialization for high performance on gpus.
\newblock In {\em Proceedings of the 19th ACM SIGPLAN symposium on Principles
  and practice of parallel programming}, pages 119--130, 2014.

\bibitem{CUTLASS}
{CUTLASS Developers}.
\newblock {Efficient GEMM in CUDA}, 2023.
\newblock
  \url{https://github.com/NVIDIA/cutlass/blob/main/media/docs/efficient_gemm.md}.

\bibitem{array}
{Silvio Ranise and Cesare Tinelli}.
\newblock {theory Arrays}, 2005.
\newblock \url{https://smtlib.cs.uiowa.edu/version1/theories/Arrays.smt}.

\end{thebibliography}
